# DYNAMIC ASSIGNMENT IN MICROSIMULATIONS OF PEDESTRIANS

**Authors:** Tobias Kretz[1], Karsten Lehmann[1,2], Ingmar Hofsäß[1], Axel Leonhardt[1]
**Affiliation**[1]: PTV Group
**Address**[1]: Haid-und-Neu-Straße 15, D-76131 Karlsruhe, Germany
**Affiliation**[2]: init AG
**Address**[2]: Käppelestraße 4-6, D-76131 Karlsruhe, Germany
**Email (corresponding author):** Tobias.Kretz@ptvgroup.com

## ABSTRACT
A generic method for dynamic assignment used with microsimulation of pedestrian dynamics is introduced. As pedestrians – unlike vehicles – do not move on a network, but on areas they in principle can choose among an infinite number of routes. To apply assignment algorithms one has to select for each OD pair a finite (realistically a small) number of relevant representatives from these routes. This geometric task is the main focus of this contribution. The main task is to find for an OD pair the relevant routes to be used with common assignment methods. The method is demonstrated for one single OD pair and exemplified with an example.

## INTRODUCTION

Finding and evaluating the user equilibrium network load for road networks (traffic assignment) is one of the core tasks of traffic planning. Various algorithms to solve this problem have been developed over the years (Wardrop, 1952) (Beckmann, McGuire, & Winsten, 1956) (LeBlanc, Morlok, & Pierskalla, 1975) (Bar-Gera, 2002) (Gentile & Nökel, 2009).

For readers who are not familiar with the concept of iterated assignment a short summary: the basic idea is to simulate (or compute) a scenario multiple times always computing the assignment on the routes based on the results of one or more or all previous simulations (iteration steps). The aim is to come up with a user-equilibirum route assignment which is –Wardrop's principle – that no driver (or pedestrian) can achieve a smaller travel time by changing the route. This implies that tall travel times of the routes for an origin destination pair are equal. The efficient computation of the assignment from previous simulation runs such that the equilibrium is reached with as few iterations as possible, is a demanding task and progress has been made throughout recent decays since Wardrop stated his principle.

The main hindrance for the application of said algorithms with pedestrian traffic is that all these algorithms are formulated for application with discrete networks, i.e. they compute flows on a node-edge graph. Pedestrians on the contrary can and do move continuously in two spatial dimensions. Their walking area can have holes (obstacles), but this does not change the fact that – even when loops are excluded – there are infinitely many possible paths for a pedestrian to walk from an origin to a destination.

There are two ways to face the resulting challenge: either one develops an entirely different assignment algorithm for pedestrian traffic – this was done for example in (Hoogendoorn & Bovy, 2004, S. a) (Hoogendoorn & Bovy, 2004, S. b) (van Wageningen-Kessels, Daamen, & Hoogendoorn, 2014)– or one extracts from the pedestrian walking area in a well-grounded suitable way a graph structure which can be used with existing assignment methods. This is what is attempted in this contribution.

In the remainder the conditions and restrictions for the application will be discussed and the task that is attempted to be accomplished in this contribution will be clearly defined. This is followed by the section which defines the main ideas of the method. Finally an example of application will be given.

## PRELUDE: NAVIGATING WITH DISTANCE MAPS

Models of pedestrian dynamics usually rest on the assumption that pedestrians predominantly move into the direction of the (spatially) shortest path toward their destination. This basic direction then is modified by other effects of which one has to be the influence of other pedestrians. For the computation of the direction of the shortest path there are a number of methods of which one is distance maps.

The method of distance maps or static potentials has been particularly popular for modeling pedestrian dynamics in particular in use with cellular automata models (Burstedde, Klauck, Schadschneider, & Zittartz, 2001) (Kirchner & Schadschneider, 2002) (Klüpfel, 2003) (Nishinari, Kirchner, Namazi, & Schadschneider, 2004) (Kretz & Schreckenberg, 2009) (Schadschneider, Klüpfel, Kretz, Rogsch, & Seyfried, 2009). The reason for this is probably that the data structure of the distance map fits well to the spatial representation of cellular automata models: a regular usually rectangular grid. FIGURE 1 visualizes an example of a distance map.

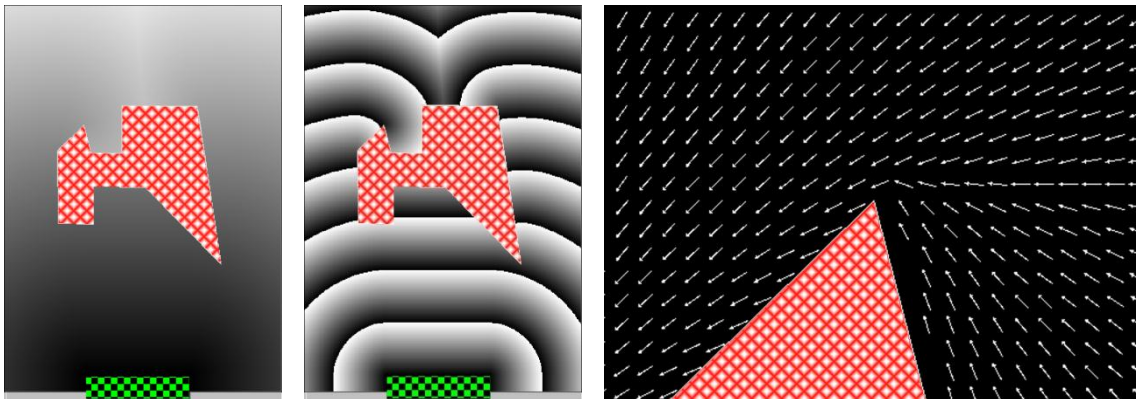

**FIGURE 1:** The left figure shows a distance map with the brightness of a spot directly proportional to the value of the grid point, i.e. the distance of that spot to the destination area. The destination area is shown as a green-black checkerboard, the obstacle is marked with diagonal red lines on white ground. In the central figure the brightness is computed modulo to some distance value $d$. The figure on the right side shows an excerpt of the geometry: shown is the area around the upper left corner of the obstacle with the field of negative gradients (white) of the distance field. It is a general mathematical property of gradients that they are orthogonal to lines of equal value (equi-potential lines) of the field from which they are derived. Therefore the negative gradients give the direction which for a given step distance most reduces the distance to destination. Following the negative gradients pedestrians are also guided around the obstacle.

In cellular automata models the distance maps are usually used such that pedestrians move with higher probability to a cell which is closer to the destination. In other modeling approaches which do not directly operate with space, but rather with velocity or acceleration as for example the force-based models do, pedestrians follow the negative gradient of the distance map at their current position. ``Follow'' here means that this direction is used as preferred or desired or base direction and may be subject to modifications as consequence of influence of for example other pedestrians or limited acceleration ability.

A standard method for the computation of the distance maps is the Fast Marching Method (Kimmel & Sethian, 1998). For other methods see (Kretz, Bönisch, & Vortisch, Comparison of Various Methods for the Calculation of the Distance Potential Field, 2010), (Schultz, Kretz, & Fricke, 2010).

# PROBLEM ILLUSTRATION

Imagine a walking geometry as shown in FIGURE 2. A pedestrian following the above mentioned principle of walking into the direction of the shortest path would pass the obstacle on the left side (reader's perspective) as it is shown in the left figure. In this case it is easy to make pedestrians pass the obstacle on either side by introducing on each side of the obstacle an intermediate destination area. With the intermediate areas and the routes which include them, it is possible to route pedestrians locally into the direction of the shortest path, but still make a given fraction of pedestrians detour: first the pedestrians would head toward one of the two intermediate destinations and as soon as it is reached proceed to the final destination.

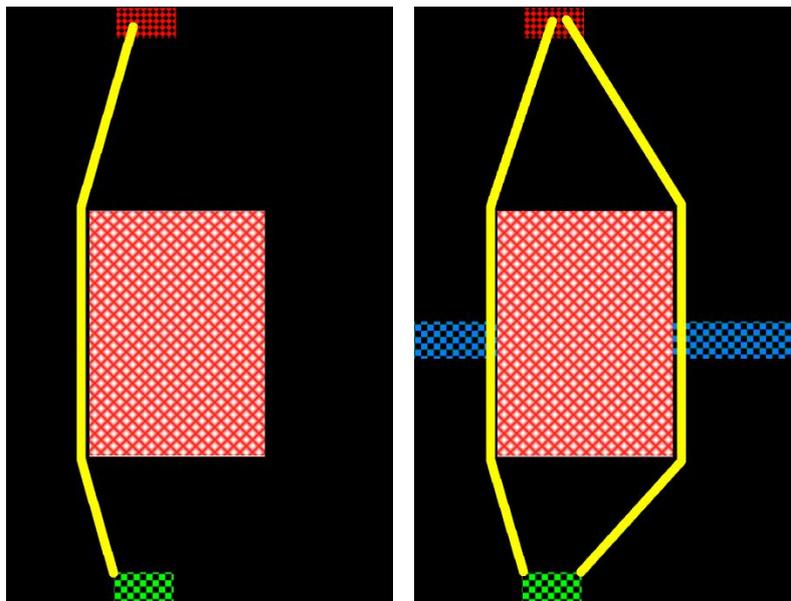

**FIGURE 2: Walking area (black), obstacle (diagonal red lines on white ground), origin area at the upper sides (red and black diagonal checkerboard pattern) and destination at the lower ends (green and black checkerboard pattern). Left figure: The yellow line shows the path a pedestrian (set into the simulation at some arbitrary coordinate on the origin area) would follow if a distance map is used to determine his basic direction. The route data simply would be some information (e.g. an area ID) identifying the destination area. Right figure: Compared to the left side figure the black-blue areas mark intermediate destination areas. There are two routes, one leading over each of the intermediate destination areas. (The intermediate destination is not necessary along the shorter (left) path, it has been added for illustration.)**

In FIGURE 2 the intermediate destination area on the left side was not necessary. Having a route directly leading from the origin to the destination area would have the same effect. However, it is important to note that the lower intermediate destination area does not change the path of the pedestrians on that route compared to the case without any intermediate destination area. This is not in general the case. In general it is difficult to shape the intermediate destinations such that the path on the principally shortest route is not distorted compared to the case without intermediate destination. FIGURE 3 shows such a case. It is concluded that the intermediate destination areas cannot be of trivial shape (rectangles) in general.

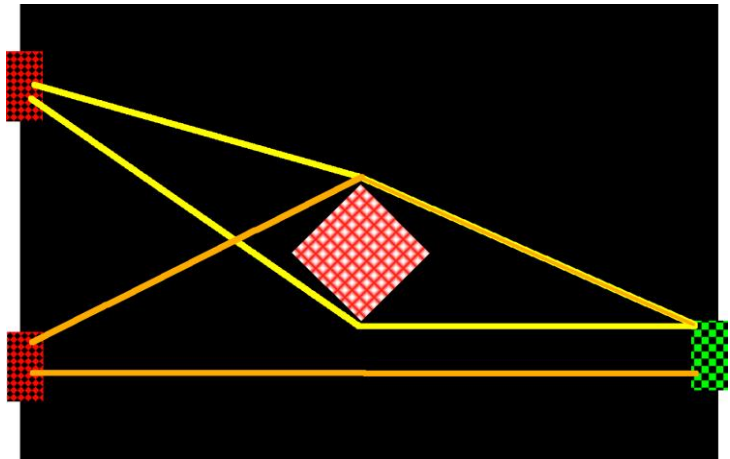

**FIGURE 3:** Example with necessarily non-trivial geometry for the intermediate destination areas. Walking area (black), obstacle (diagonal red stripes on white ground), two origin areas to the left (red-black diagonal checkerboard), destination area to the right (green-black checkerboard), and shortest paths (yellow and orange). Compare FIGURE 4.

The basic idea is now that the intermediate destination areas need to be shaped along the equi-distance (equi-potential) lines of the destination area distance map (static potential). This is shown in FIGURE 5.

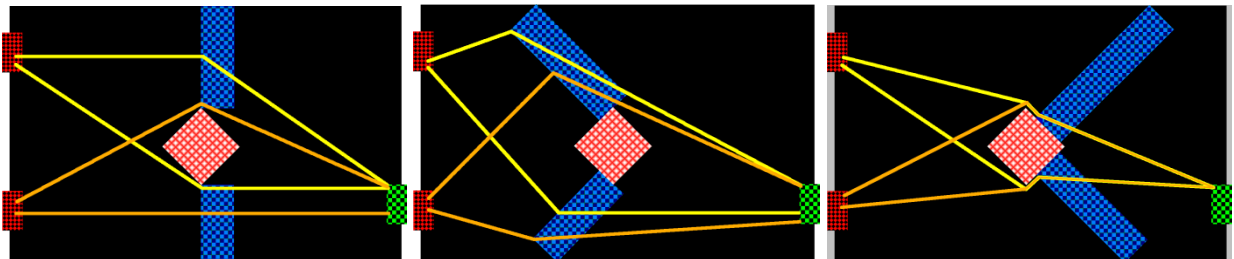

**FIGURE 4:** For the example of FIGURE 3 simple rectangular areas are used as intermediate destination areas (light and dark blue checkerboard pattern). With none of the three variants the shortest paths as shown in FIGURE 3 are reproduced.

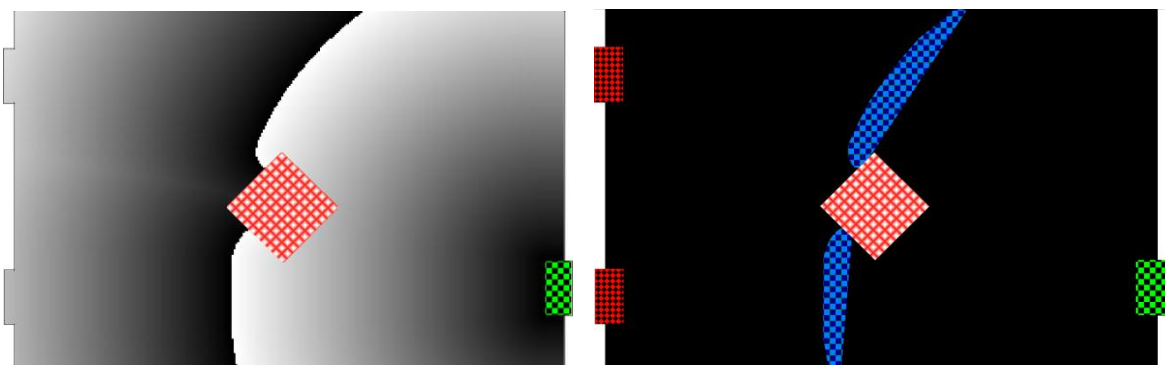

**FIGURE 5:** The left side shows the distance map of the destination area. Brighter pixels show larger distance. However, about at half the maximum distance the brightness is set back to zero to better display one distinct equi-distance line. The right side shows the geometry with two intermediate destination areas added which are shaped along that particular equi-distance line with their left side. The exact shape of the right side is not relevant. With these intermediate destination

**areas the shortest paths of FIGURE 3 are reproduced necessarily and exactly up to the limited precision in the numerical definition of the intermediate destination areas.}**

# METHOD FORMULATION

In this section a method will be formulated that computes the geometry of intermediate destination areas for routes which represent the main routing alternatives for pedestrians.

A note in advance: the method will be proposed here only insofar it is necessary for the concluding example of application. The complete method includes two more elements to deal with certain more difficult geometries. As a consequence of the strictly limited extent of this publication we have omitted these parts. The publication of the full method is currently in preparation (Kretz, Lehmann, & Hofsäss, 2014). The method will be defined and explained visually and not in a mathematical language.

The first step is to compute a distance map for the destination (respectively for all destinations; this option is implied in the following) for which alternative routes ought to be computed. Next the modulo variant of this potential as it is displayed in the middle of FIGURE 1 is computed. This is not only a useful visualization to see the properties of the distance map clearly, but its structure is also an important element of the method. FIGURE 6 shows for a simple example geometry the same potential for various modulo values *d*.

FIGURE 7 shows a modification of FIGURE 6. The gradual color change from black to white has been omitted. Instead one range from black to white is shown in one and the same color and the next range in a different one. In this way areas which are within a range of distances to destination are grouped.

The coloring of FIGURE 7 (left and center figure) is the key for the further steps in the method. To be precise: all regions which share the property of the magenta colored region have to be identified. These are regions, which are unique for their range of distances to destination, but which have two unconnected regions directly neighbored which are closer to the destination than they are.

Relevant origin positions of a pedestrian in a simulation of the scenario of FIGURE 7 are the light and dark orange and the magenta regions as well as the light and dark cyan regions which are further away from the destination than the magenta area. If a pedestrian's origin position is on a cyan area which is closer to the destination than the closest orange area no route alternatives are generated.

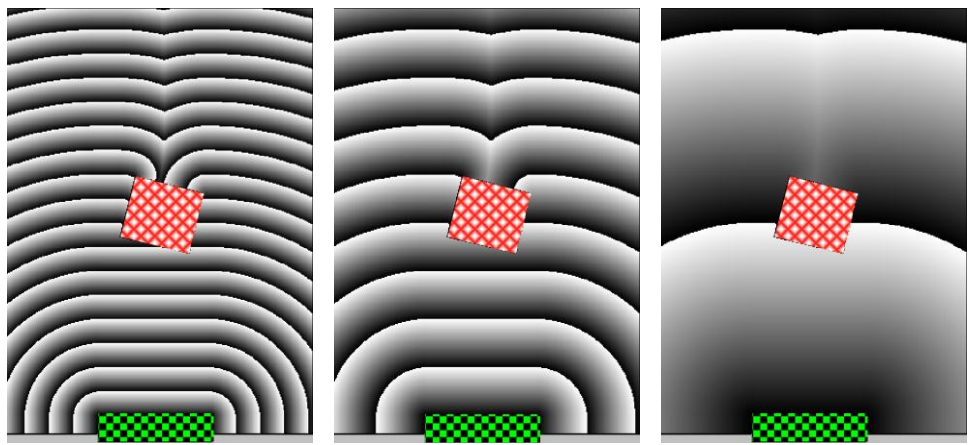

FIGURE 6: One potential displayed with three different modulo values *d*. The ratio of the modulo values of the middle and the left display is 2, the one of the right vs. the left one is 8. The destination is colored as black and green checkerboard, diagonal red lines on white ground depict an obstacle.

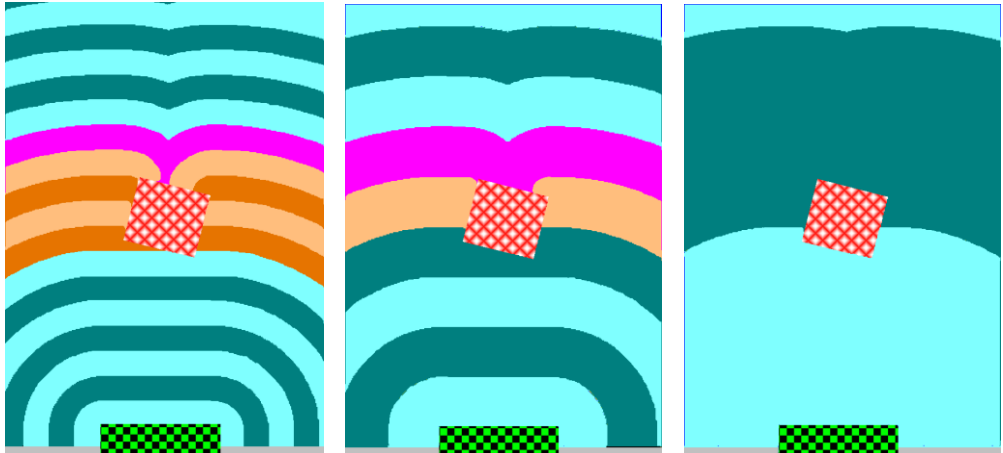

FIGURE 7: The destination is shown as green and black checkerboard, an obstacle is colored with diagonal red lines on white ground. Light and dark cyan mark regions which are simply connected and which lie within a range of distances to destination and which have exactly one simply connected area as direct neighbor which is closer to destination. Light and dark orange mark regions where there is more than one (here: exactly two) unconnected (split by an obstacle) regions which are within a range of distances to destination. The magenta region is simply connected and has more than one (here: exactly two) directly neighboring areas which are closer to destination than the magenta region is itself.

In almost all simulation models of pedestrian dynamics the base direction is set into the direction of the shortest path. This implies that it is determined by the origin position if a pedestrian passes the (red) obstacle to the right or to the left side. If we want to choose freely if a pedestrian passes the obstacle either to the left or the right side, the magenta region in FIGURE 7 is the critical one. Once a pedestrian has passed it, the shortest path paradigm will take the pedestrian to the destination.

Therefore the route choice if the obstacle is passed to the left or to the right side is modeled by making the two light orange regions immediately neighbored to the magenta region intermediate destination areas. Each of the two is assigned to one route, generating a route choice.

Obviously not only the shape of the additional intermediate destination areas depends on parameter *d*, but first and foremost if it exists at all. The right figure of FIGURE 7 shows no additional intermediate destination areas because the value of parameter *d* has been chosen too large

The choice of the value of parameter *d* at first might appear to be arbitrary and its existence therefore as a major downside of the method. However, simulation models of pedestrian dynamics usually contain elements to deal with obstacles. These elements work the better -- i.e. they capture a larger share of the effect of the obstacle -- the smaller the obstacle is. Therefore it is not desired to have a litter box create a route alternative, but one wants such small obstacles be handled by the operational pedestrian dynamics model and the value of *d* should be chosen at least that large that such small objects do not create route alternatives. Second, for the purpose of computational efficiency and limited computation resources, in large simulation projects it can be desirable to not consider every existing routing alternative, but only the major ones. Then one may set the value of parameter *d* to even larger values to maintain a feasible simulation project size.

The left figure of FIGURE 7 suggests that there is some arbitrariness in selecting the regions which are directly neighbored to the magenta region as intermediate destination areas. Why not two of the dark orange or the other two light orange regions? The answer can be found by considering pedestrians who have an origin position somewhere on the dark orange or light orange regions. It is desired to be able to also assign such pedestrians a route which leads over an intermediate destination area on the opposite side of the obstacle. If that intermediate destination area was moved toward the destination, some pedestrians would approach the intermediate destination from the back side. This implies an unrealistic trajectory and is therefore not desired. In fact because any intermediate destination area has a finite depth, there can be a small area where this occurs although the region closest to the magenta region has been made an intermediate destination area. This problematic region is the larger, the larger the value of parameter $d$ is. Realizing this it is important to realize as well that only the shape of the front side of an intermediate destination is relevant. The depth can be reduced to the point that it is still guaranteed that any pedestrian heading to the intermediate destination does not have to slow down to step on it for at least one time step. Nevertheless for the remainder of this contribution we will not modify the shape of intermediate destination areas in this way. The unmodified shape simply gives a clearer impression of the idea.

**Multiple Obstacles**

If there is more than one obstacle in a scenario the idea presented so far has to be applied recursively. This is shown in FIGURE 8. The left and middle figure show the initial computation step for and with the potential. The dark orange regions are identified as intermediate destinations to pass the downstream obstacle, the light orange regions have the same role for the upstream obstacle. However, the shape of the upstream regions is computed with respect to the destination. If we want to be able to send pedestrians without artifacts in their trajectories on routes which lead from an upstream intermediate destination area to that downstream intermediate area which is more remote respectively, we have to calculate the shape of the upstream intermediate destination areas with respect to the downstream intermediate destination areas. For one of the two downstream intermediate destination areas this is demonstrated in the figure on the right side of FIGURE 8.

The brown region in FIGURE 8 (right side figure) -- the region which is closer to the destination than the downstream intermediate destination area -- does not only mark the region where pedestrians' origin positions do not generate routing alternatives. When the map of distances -- the potential -- is computed beginning at the downstream intermediate destination area, the brown region must not be intruded. It is therefore treated as if it was an obstacle -- for the computation of routing alternatives, not later in the simulation. If the brown region was intruded during the computation of the potential short cut paths could result which would lead to unrealistic movement and thus spoil the intended effect of the method.

FIGURE 9 finally shows all resulting intermediate destination areas and resulting routes. One can see there that the original route without intermediate destinations is preserved -- it is the route which ends at the central of the nine destination dots which are depicted in green. Of the nine routes the one to the first green destination point from bottom, third from bottom and fifth from bottom basically are identical as they either have no intermediate destination point or if they have they are along the globally shortest path. Does this imply a problem or an inefficiency for the usage of these routes in an assignment method? It is surely not a serious problem. Most assignment methods can deal with multiple identical routes. In realistic networks there are always routes which overlap. Assignment methods have to cope with this. Identical routes are simply routes which overlap by 100%. Second, we have to note that the answer on the question which routes are identical depends on the location of the starting area. If the starting area in FIGURE 9 were in the top corner the route without intermediate destinations were identical to different routes. It may even happen that this property is different for different sections of a starting area. This makes clear that only an additional a posteriori analysis under consideration of the starting area can show which routes could be dropped.

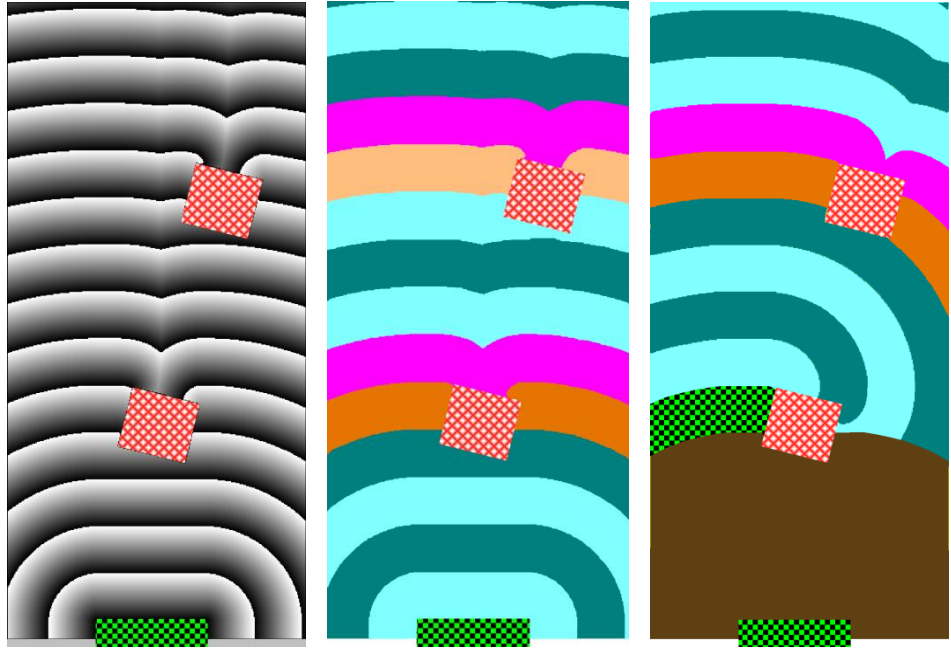

**FIGURE 8:** Compared to the previous example a second obstacle has been added. Right side figure: Computation of the upstream intermediate destination areas with respect to one of the downstream intermediate destination areas. The region where pedestrians' origin positions do not generate route alternatives is shown brown. It is a virtual obstacle into which in the further recursions of the algorithm not anymore a potential will spread.

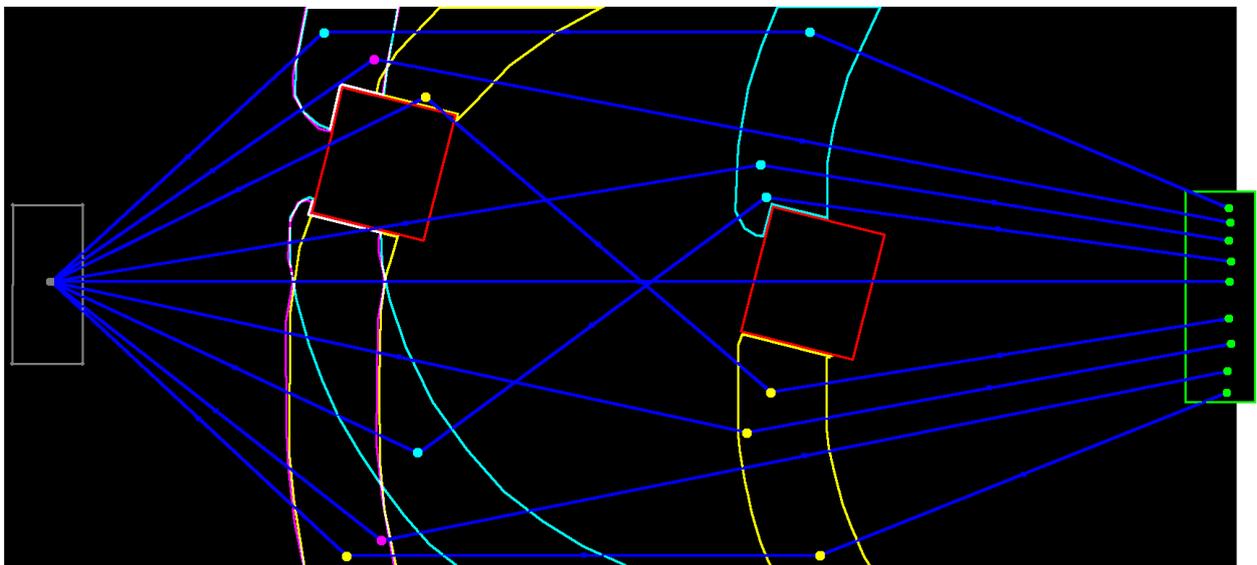

**FIGURE 9:** Shown here are the edge of the destination area (green), the edges of the obstacles (red), the edge of some origin area (gray), the edges of the downstream intermediate destination areas (cyan and yellow) and the upstream intermediate destination areas (cyan, yellow, magenta). Upstream and downstream intermediate destination areas which follow in a sequence on one route are shown in the same color (cyan or yellow). Furthermore this figure shows the routes

in blue. The origin area is marked with a gray dot, the destination area with green dots. All intermediate destination areas are marked with the color that corresponds to the color of the edge of the area where they are attached. The dots only mark the area which is an intermediate destination area. Their position on the area has no impact on the movement of pedestrians in the simulation.

**Nested route choices**

FIGURE 9 raises the question, if the intermediate destination areas whose edges are colored magenta could be dropped together with their routes. Obviously the two routes do not actually add an alternative path choice. So is it sufficient to always just consider that routing alternative which is currently closest to destination (in FIGURE 8 marked by the dark orange areas in the lower figure) for further process and omit any more remote one (the light orange areas in FIGURE 8)?

The answer is``no'' and the reason for this is that there are cases where route alternatives are nested; or in other words: it cannot necessarily be decided what is ``upstream'' and what ``downstream''. FIGURE 10 demonstrates this with an example. It can -- of course -- always be decided which of two points is spatially closer to a destination. However, in terms of travel time this is different. One can easily imagine in the third and fourth figure of FIGURE 10 where there has to be a jamming crowd of pedestrians to make the indicated path the quickest to destination although it implies and includes a spatial detour.

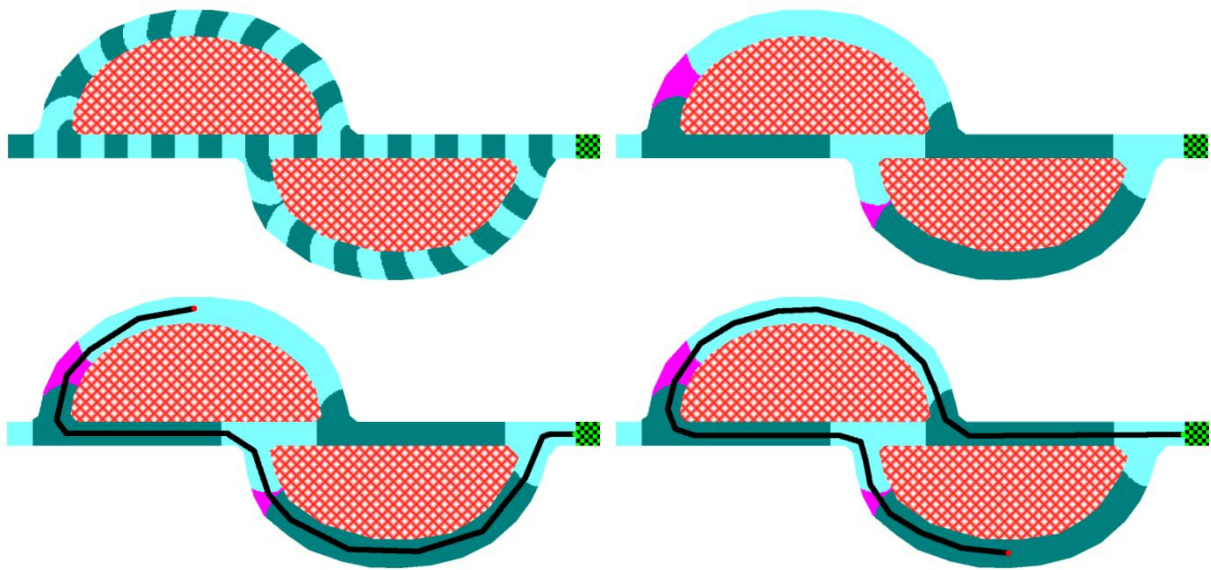

FIGURE 10: In this example the route choices are nested. The destination is on the right side. The first computation of the potential (upper left figure) brings up two route alternatives, respectively two local maxima of distance to destination. These are manifest by the existence of the two regions marked with magenta in the upper right figure. lower figures show two different possible trajectories (black; from red to green) of pedestrians where the two magenta regions are passed in different sequence.

## EXAMPLE APPLICATION

The geometry of FIGURE 10 but with two additional bottlenecks on the straight connection between origin and destination is used for a demonstration, see FIGURE 11 for details. With this geometry, these routes and a demand of 5.000 pedestrians per hour an iterated assignment is carried out using PTV Viswalk (Kretz, Hengst, & Vortisch, 2008) for all operational aspects. This is by no means required as the method is very generic and can be combined with various operational models of pedestrian dynamics.

The geometry of this example is comparatively close to a node-link network. Therefore the route choices can relatively easy be determined "by hand" without the help of the proposed algorithm. Therefore the choice of this geometry may appear to be not a good coice. However, if one is able to reproduce the results of the algorithm without any tool, one can confirm that these results make sense. Other examples with a different focus are published elsewhere (Kretz, Lehmann, & Friderich, 2013) (Kretz, Lehmann, & Hofsäss, 2013).

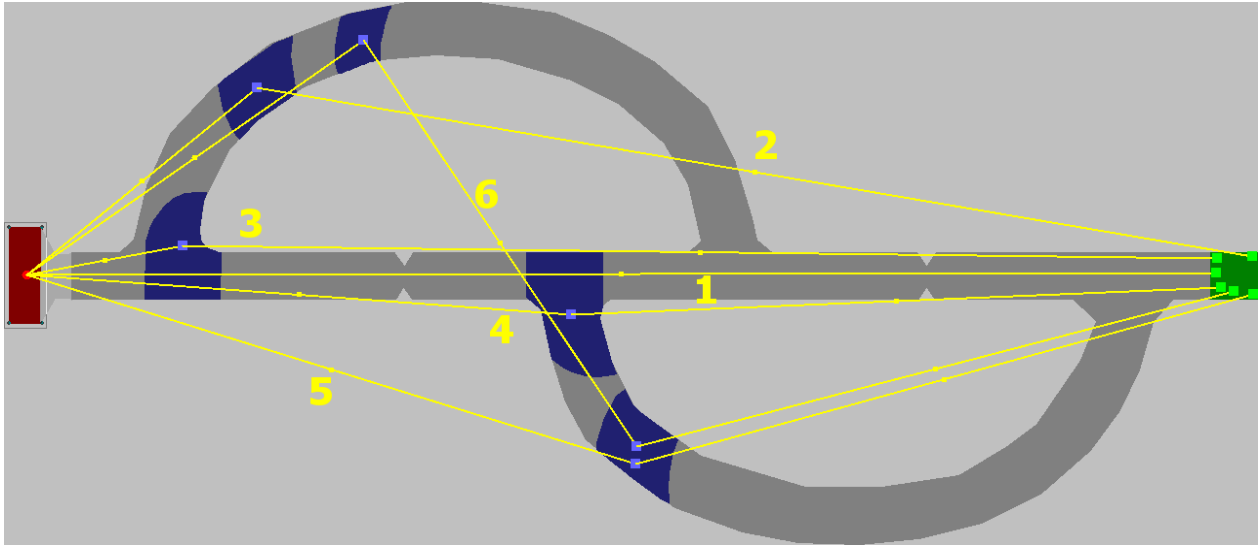

FIGURE 11: Example geometry with calculated intermediate destinations (blue) and routes with their IDs (yellow). Note the bottlenecks along the straight connection between origin (red) and destination (green). From left to right the model extends over 106 m. The corridors are 4 m wide, respectively 2 m at the two bottlenecks. Note that there are two different intermediate destinations on the upper arc because one is computed based on the final destination and one based on the intermediate destination on the lower arc.

As initial distribution we've set for once a heavy load on the routes along the globally shortest connection (routes 1, 3, and 4) and in a second run an equal distribution on all routes. After each iteration step, the routes with the maximum and the minimum average travel time are selected. Based on the number of possible routes and the time difference between the maximum and minimum average travel time the load (i.e. the probability for a pedestrian to choose a route) on the routes with maximum travel time is decreased and the load on routes with the minimum travel time is increased by the same amount, which is calculated as

$$\left(\frac{tMax - tMin}{tMax + tMin}\right)^{\delta}$$

where $\delta$ is a free parameter, which has been set to $\delta=1.0$ for this example. To avoid oscillations a further reduction factor has been applied if from one iteration to the other the route with smallest travel time has become the route with longest and the route with longest travel time has become the route with the smallest travel time. We do not elaborate on these details because the geometric aspects are in the focus of this paper. For the same reason we have chosen this simple assignment algorithm which only makes use of the results of the immediately previous iteration's results instead of making use of all results.

As a side note: in reality pedestrians do not choose their route exclusively based on travel time. One would rather apply a generalized cost approach. With generalized costs it is possible – at least in principle

– to consider other aspects as ground surface, light conditions, social aspects or landmarks which presumably lead to observed route choice behavior which obviously are not travel time based, at least not exclusively (Graessle & Kretz, 2011).We stick here with the simpler idea to base route choice purely on travel time as the geometrical aspects are the focus of the facing work and a discussion on the details of the generalized costs would clearly exceed the scope and available extent of this contribution. It needs furthermore be clear that the results of an iterated assignment approach are not connected to local conditions at spots and times when pedestrians in reality can and do choose about a route or a part of a route.

Routes with a value of zero get a small load during simulation to take them into account if they get effective again in later iteration steps. The average travel time of these routes is only taken into account if it is the minimum average travel time. The iteration is continued until either only one route remains, the average travel time on all routes differ only by a predefined value from each other or the number of iterations exceeds a predetermined value, which was set here to half a second.

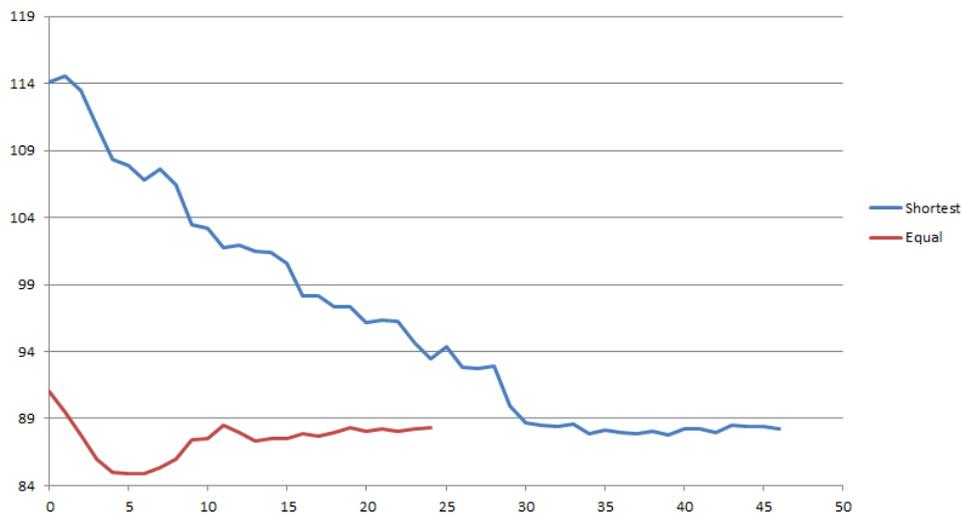

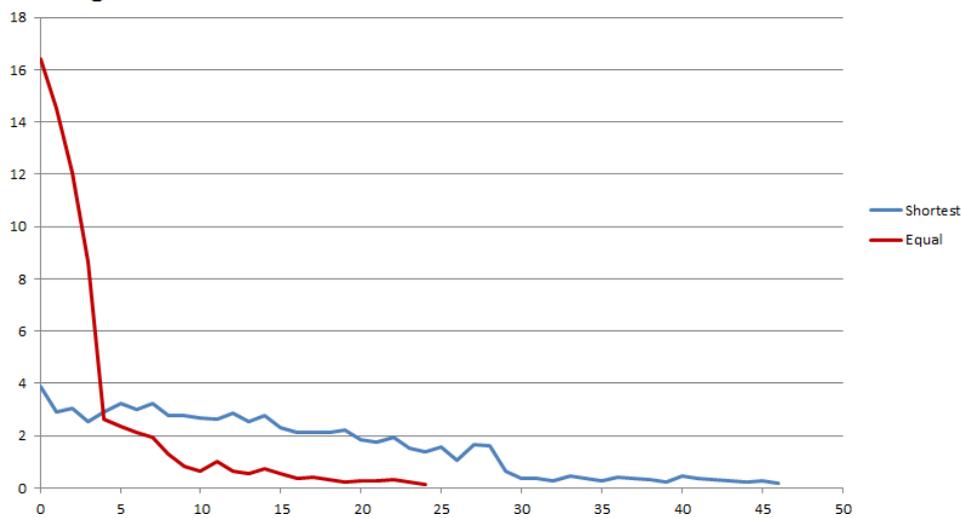

**FIGURE 12:** The upper figure shows the weighted average travel time and the lower figure the weighted standard deviations of travel times evolve in the course of iterations. The blue line shows the evolution when the initial distribution is mainly on the globally shortest path and only one percent is assigned to each of the three detouring routes. The red line shows the evolution when the initial distribution is equal on all routes.

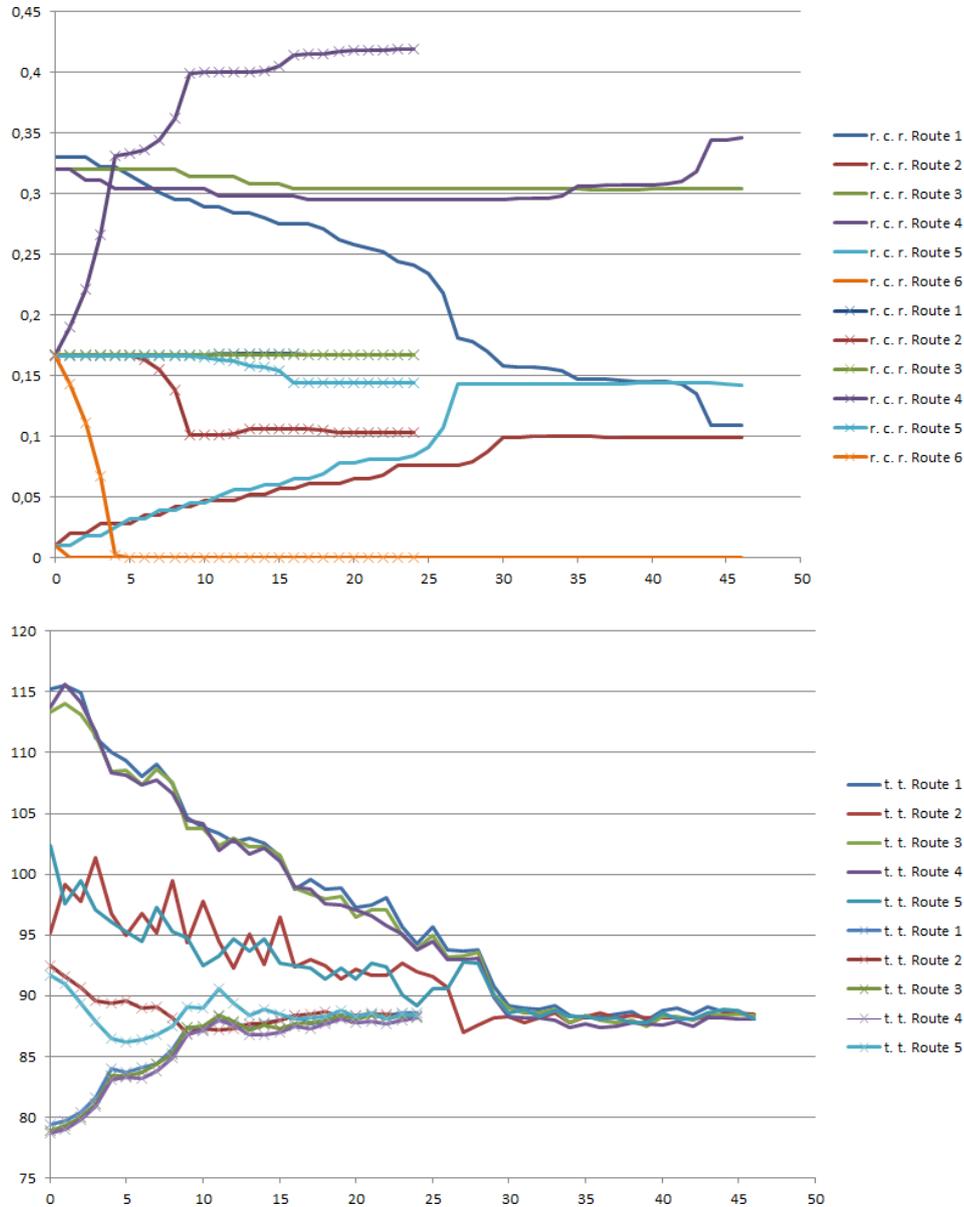

**FIGURE 13:** The upper figure shows the evolution of route choice ratios for each route and for both cases of initial distribution (lines for equal distribution are marked with a cross). The figure on the lower right shows the evolution of travel times for each route over the course of iterations.

FIGURE 13 and FIGURE 13: The upper figure shows the evolution of route choice ratios for each route and for both cases of initial distribution (lines for equal distribution are marked with a cross). The figure on the lower right shows the evolution of travel times for each route over the course of iterations.show the results for an iterated assignment in the geometry of FIGURE 11 with a demand of 16,000 pedestrians per hour. Travel times for all pedestrians arriving at the destination in the time interval 300..900 sec were considered. It can be seen that the total average travel time at equilibrium is the same (88.3 sec) independent of the route choice ratios at the first iteration. Similar holds for the route choice ratios of the three detouring routes 2 (9.9% resp. 10.3%), 5 (14.2% resp. 14.4%) and 6 (0%). However the three routes 1, 3, and 4, which all lead along the globally shortest path, end up with largely different route choice ratios for different ratios in the initial iteration. However, in sum the route choice ratios do not depend on the initial ratios. This is shown in FIGURE 14 where the sum of route choice ratios and the average of travel times of route 1, 3, and 4 are shown next to the ones of routes 2, 5, and 6.

That the route choice ratios on routes 1, 3 and 4 end up differently for different initial values while the sum is the same is a clear indication that the introduction of the intermediate destination does not have a significant effect on local movement behavior. Because if it had, the travel times on these three routes actually would be different and the route choice ratios for the three routes should be mutually different, but each of them should be identical for different initial ratios. For only if the travel times are identical the route choice ratios cannot be determined uniquely.

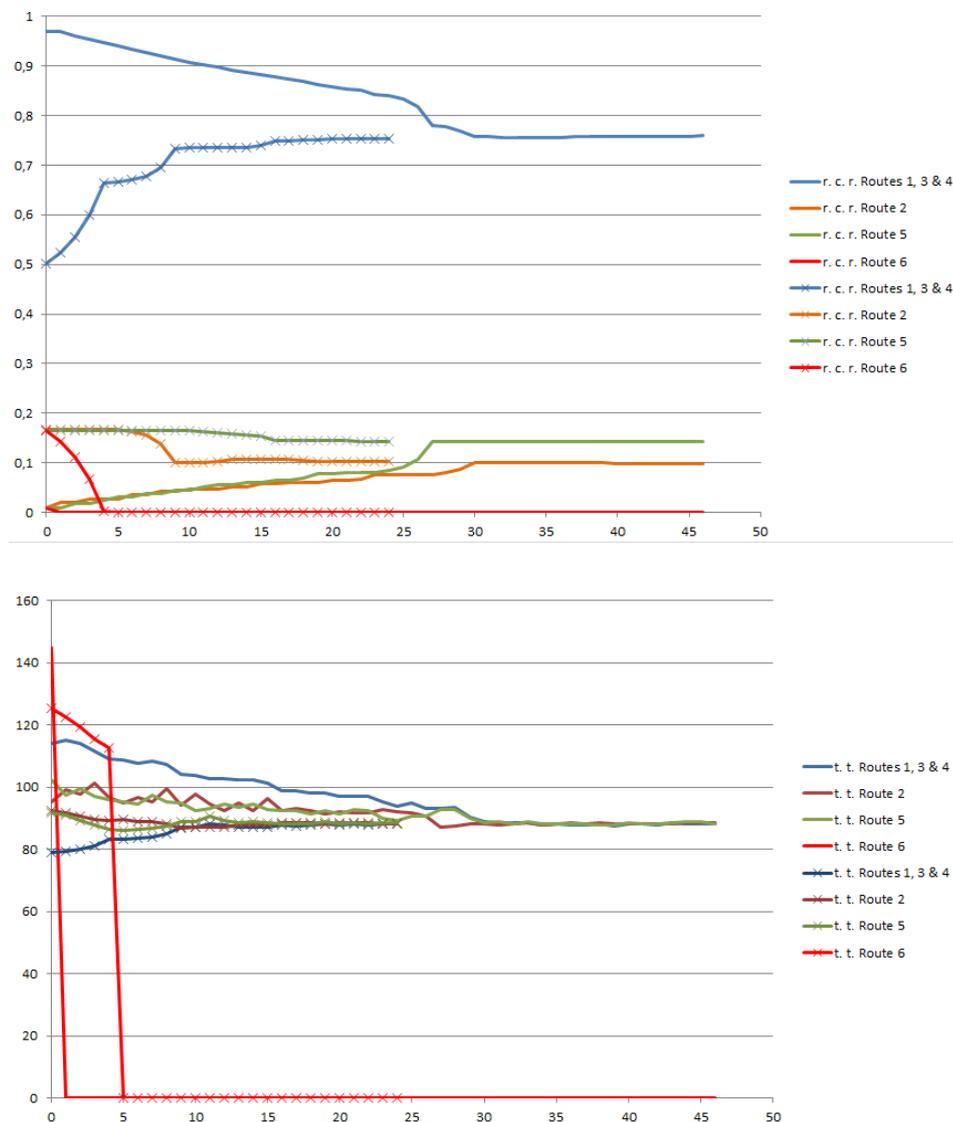

**FIGURE 14: Route choice ratio (top) and travel times (bottom) with routes 1, 3, and 4 averaged or added, respectively.**

# CONCLUSIONS

We have sketched the core of an algorithm that produces routes with intermediate destination areas to be applied with pedestrian simulation. The routes which are computed mark the most relevant routing alternatives in any given walking geometry. Having reduced the infinitely many trajectories by which a pedestrian can move from origin to destination to a small set of routes it is possible to apply iterated assignment methods to compute the user equilibrium of any given scenario (walking geometry and OD matrix). We have demonstrated the capability of the approach with an example.

In our opinion it is less questionable that the method can produce good results in arbitrary scenarios in principle but it rather remains to be shown that the method is computationally feasible also for large scale, real life planning work.

The proposed method of iterated assignment is not meant to replace earlier one-shot assignment methods (Pettré, Grillon, & Thalmann, 2007) (Kirik, Yurgel'yan, & Krouglov, 2009) (Kretz, Pedestrian Traffic: on the Quickest Path, 2009) (Guy, et al., 2010), (Kretz, et al., 2011), (Kemloh Wagoum, Seyfried, & Holl, 2012). One-shot assignment has to build on current information, be it global or local (only what a pedestrian can perceive in his environment) and instantaneous or with a delay or extrapolation into the near future. Iterated assignment and the idea of a dynamic equilibrium on the other hand rest on the assumption that pedestrians have information of the load of the walking geometry (or road network) throughout the full time of their presence in the system. This includes – for the early phase of their trip – knowledge of a relatively far future. This can only be justified in situations which repeat similarly on a regular base. Obviously both modeling approaches have their justification for shares of all possible planning problems. The question is how to handle scenarios which are a combination of both. One can easily imagine that scenarios are repeating only approximately and therefore one has to imagine the system to be near equilibrium but with large shares of pedestrians actively and continuously searching for better options based on what they perceive currently. The problem is that one-shot and iterated assignment are not easily compatible. Thus this remains a puzzle for the future. However, we can very well imagine that the geometric part of this contribution can be combined with a one-shot assignment approach.

# LITERATURE

Bar-Gera, H. (2002). Origin-based algorithm for the traffic assignment problem. *Transportation Science*, pp. 398-417. DOI: 10.1287/trsc.36.4.398.549

Beckmann, M., McGuire, C., & Winsten, C. (1956). *Studies in the Economics of Transportation.* Yale University Press.

Burstedde, C., Klauck, K., Schadschneider, A., & Zittartz, J. (2001). Simulation of pedestrian dynamics using a two-dimensional cellular automaton. *Physica A*, pp. 507-525. DOI: 10.1016/S0378-4371(01)00141-8 arxiv.org/abs/cond-mat/0102397

Gentile, G., & Nökel, K. (2009). Linear User Cost Equilibrium: the new algorithm for traffic assignment in VISUM. *Proceedings of European Transport Conference 2009*, p. eprint.

Graessle, F., & Kretz, T. (2011). An Example of Complex Pedestrian Route Choice. *Pedestrian and Evacuation Dynamics* (pp. 767--771). Gaithersburg: Springer. DOI: 10.1007/978-1-4419-9725-8_71 arxiv.org/abs/1001.4047

Guy, S., Chhugani, J., Curtis, S., Dubey, P., Lin, M., & Manocha, D. (2010). Pledestrians: A least-effort approach to crowd simulation. *SIGGRAPH 2010* (pp. 119-128). Eurographics Association. DOI: 10.2312/SCA/SCA10/119-128

Hoogendoorn, S., & Bovy, P. (2004). Dynamic user-optimal assignment in continuous time and space. *Transportation Research Part B: Methodological*, pp. 571-592. DOI: 10.1016/j.trb.2002.12.001

Hoogendoorn, S., & Bovy, P. (2004). Pedestrian route-choice and activity scheduling theory and models. *Transportation Research Part B: Methodological*, pp. 169-190. DOI: 10.1016/S0191-2615(03)00007-9

Kemloh Wagoum, A., Seyfried, A., & Holl, S. (2012). Modeling the dynamic route choice of pedestrians to assess the criticality of building evacuation. *Advances in Complex Systems, 15*(7). DOI: 10.1142/S0219525912500294 arxiv.org/abs/1103.4080

Kimmel, R., & Sethian, J. (1998). Computing geodesic paths on manifolds. *PNAS, 95*(15), pp. 8431--8435.

Kirchner, A., & Schadschneider, A. (2002). Simulation of evacuation processes using a bionics-inspired cellular automaton model for pedestrian dynamics. *Physica A*, pp. 260-276. DOI: 10.1016/S0378-4371(02)00857-9 arxiv.org/abs/cond-mat/0203461

Kirik, E., Yurgel'yan, T., & Krouglov, D. (2009). The shortest time and/or the shortest path strategies in a CA FF pedestrian dynamics model. *Journal of Siberian Federal University. Mathematics & Physics, 2*(3), pp. 271-278.

Klüpfel, H. (2003). *A Cellular Automaton Model for Crowd Movement and Egress Simulation.* Duisburg: Universität Duisburg-Essen.

Kretz, T. (2009). Pedestrian Traffic: on the Quickest Path. *Journal of Statistical Mechanics: Theory and Experiment*, p. P03012. DOI: 10.1088/1742-5468/2009/03/P03012 arxiv.org/abs/0901.0170

Kretz, T., & Schreckenberg, M. (2009). F.A.S.T. - Floor field- and Agent-based Simulation Tool. In E. Chung, & A.-G. Dumont, *Transport simulation: Beyond traditional approaches.* Lausanne: EPFL press. arxiv.org/abs/physics/0609097

Kretz, T., Bönisch, C., & Vortisch, P. (2010). Comparison of Various Methods for the Calculation of the Distance Potential Field. *Pedestrian and Evacuation Dynamics 2008*, (pp. 335-346). DOI: 10.1007/978-3-642-04504-2_29 arxiv.org/abs/0804.3868

Kretz, T., Große, A., Hengst, S., Kautzsch, L., Pohlmann, A., & Vortisch, P. (2011). Quickest Paths in Simulations of Pedestrians. *Advances in Complex Systems, 14*(5), pp. 733-759. DOI: 10.1142/S0219525911003281 arxiv.org/abs/1107.2004

Kretz, T., Hengst, S., & Vortisch, P. (2008). Pedestrian Flow at Bottlenecks-Validation and Calibration of Vissim's Social Force Model of Pedestrian Traffic and its Empirical Foundations. *8th International Symposium on Transport Simulation* (p. on CD). Surfer's Paradise, Queensland, Australia: Monash University. arxiv.org/abs/0805.1788

Kretz, T., Lehmann, K., & Friderich, T. (2013). Selected Applications of a Dynamic Assignment Method for Microscopic Simulation of Pedestrians. *European Transport Conference.* Association for European Transport. abstracts.aetransport.org/paper/index/id/125/confid/1

Kretz, T., Lehmann, K., & Hofsäss, I. (2013). Pedestrian Route Choice by Iterated Equilibrium Search. *Traffic and granular Flow.* Springer (to be published).

Kretz, T., Lehmann, K., & Hofsäss, I. (2014). User Equilibrium Route Assignment for Microscopic Pedestrian Simulation. (in review). arxiv.org/abs/1401.0799

LeBlanc, L., Morlok, E., & Pierskalla, W. (1975). An efficient approach to solving the road network equilibrium traffic assignment problem. *Transportation Research*, pp. 309-318. DOI: 10.1016/0041-1647(75)90030-1

Nishinari, K., Kirchner, A., Namazi, A., & Schadschneider, A. (2004). Extended Floor Field CA Model for Evacuation Dynamics. *IEICE Trans. Inf. & Syst.*, pp. 726-732. arxiv.org/abs/cond-mat/0306262

Pettré, J., Grillon, H., & Thalmann, D. (2007). Crowds of moving objects: Navigation planning and simulation. *IEEE International Conference on Robotics and Automation*, (pp. 3062-3067).

Schadschneider, A., Klüpfel, H., Kretz, T., Rogsch, C., & Seyfried, A. (2009). Fundamentals of Pedestrian and Evacuation Dynamics. In A. Bazzan, & F. Klügl (Eds.), *Multi-Agent Systems for Traffic and Transportation Engineering* (pp. 124-154). Hershey, PA, USA: Information Science Reference. DOI: 10.4018/978-1-60566-226-8.ch006

Schultz, M., Kretz, T., & Fricke, H. (2010). Solving the Direction Field for Discrete Agent Motion. *Cellular Automata - 9th International Conference on Cellular Automata for Research and Industry, ACRI 2010*, (pp. 489-495). DOI: 10.1007/978-3-642-15979-4_52 arxiv.org/abs/1008.3990

van Wageningen-Kessels, F., Daamen, W., & Hoogendoorn, S. (2014). Pedestrian Evacuation Optimization - Dynamic Programming in Continuous Time and Space. In M. Boltes, M. Chraibi, A. Schadschneider, & A. Seyfried, *Traffic and Granular Flow 13*. Jülich: Springer.

Wardrop, J. (1952). Road Paper. Some theoretical aspects of road traffic research. *ICE Proceedings: Engineering Divisions*, (pp. 325-362). DOI: 10.1680/ipeds.1952.11259